\documentclass[aps,prb,twocolumn,showpacs,preprintnumbers,amsmath,amssymb,groupedaddress,noshowkeys]{revtex4}
\usepackage{txfonts}
\usepackage[dvipdfm]{graphicx}
\usepackage{bm}
\def\rnum#1{\expandafter{\romannumeral #1}} 
\def\Rnum#1{\uppercase\expandafter{\romannumeral #1}}

\newfont{\bg}{cmr10 scaled\magstep4}

\newcommand{\bigzerou}{\smash{\lower1.8ex\hbox{\bg 0}}}

\begin{document}

\title{Critical current oscillation by magnetic field in
semiconductor nanowire Josephson junction}
\author{Tomohiro Yokoyama}
\email[E-mail me at: ]{T.Yokoyama@tudelft.nl}
\affiliation{Kavli Institute of Nanoscience, Delft University of Technology,
Delft, The Netherlands}
\affiliation{Center for Emergent Matter Science, RIKEN Institute, Wako, Japan}

\author{Mikio Eto}
\affiliation{Faculty of Science and Technology, Keio University, Yokohama, Japan}

\author{Yuli V.\ Nazarov}
\affiliation{Kavli Institute of Nanoscience, Delft University of Technology,
Delft, The Netherlands}
\date{\today}

\begin{abstract}
We study theoretically the critical current in semiconductor nanowire
Josephson junction with strong spin-orbit interaction.
The critical current oscillates by an external magnetic field.
We reveal that the oscillation of critical current depends on
the orientation of magnetic field in the presence of spin-orbit interaction.
We perform a numerical simulation for the nanowire by using
a tight-binding model. The Andreev levels are calculated as
a function of phase difference $\varphi$ between two superconductors.
The DC Josephson current is evaluated from the Andreev levels in
the case of short junctions. The spin-orbit interaction induces
the effective magnetic field. When the external field is parallel with
the effective one, the critical current oscillates accompanying
the $0$-$\pi$ like transition. The period of oscillation is longer as
the angle between the external and effective fields is larger.
\end{abstract}
%\preprint
\maketitle

\section{Introduction}
The spin-orbit (SO) interaction has attracted a lot of interest.
In narrow-gap semiconductors, such as InAs and InSb, 
the strong SO interaction has been reported and many phenomena
based on the SO interaction are investigated intensively, e.g.,
spin Hall effect~\cite{Kato}.
The SO interaction has a great advantage also for the application to
the spintronic devices and to quantum information processing.
InAs and InSb nanowires are interesting platform for
the application and studied in recent experiments, e.g.,
the electrical manipulation of single electron spins in
quantum dots fabricated on the nanowires~\cite{Nadj-Perge}.
Nanowire-superconductor hybrid systems have been also
examined for the search of Majorana fermions induced by
the SO interaction and the Zeeman effect~\cite{Mourik}.

In Josephson junctions, the supercurrent flows when the phase
difference $\varphi$ between two superconductors is present.
In this paper, we investigate theoretically the Josephson junction of
semiconductor nanowires with strong SO interaction.
The supercurrent in semiconductor nanowires has been reported
by experiment groups~\cite{Doh,private}. The Josephson effect with
SO interaction has been studied theoretically for some materials,
e.g., magnetic normal metals~\cite{Buzdin}, where the combination of
SO interaction and exchange interaction results in an unconventional
current-phase relation, $I (\varphi) = I_0 \sin (\varphi - \varphi_0)$.
The phase shift $\varphi_0$ deviates the ground state of junction
from $\varphi = 0$ or $\pi$, which is so-called $\varphi_0$-state.
The anomalous supercurrent is obtained at $\varphi =0$.
In previous studies, we have pointed out that the anomalous effect
is attributed to the spin-dependent channel mixing due to the SO
interaction~\cite{YEN1,YEN2}. In the present study, we focus on
the critical current oscillation when an external magnetic field is applied.
The DC Josephson current is evaluated from the Andreev levels in
the case of short junction. We examine a numerical calculation using
a tight-binding model for the nanowire. In this model,
a particular form of SO interaction can be considered.
When the angle between the external field and
an effective magnetic field due to the SO interaction is smaller,
the oscillation period of critical current is shorter.

\section{Model}

\begin{figure}[t]
\includegraphics[width=6.5cm]{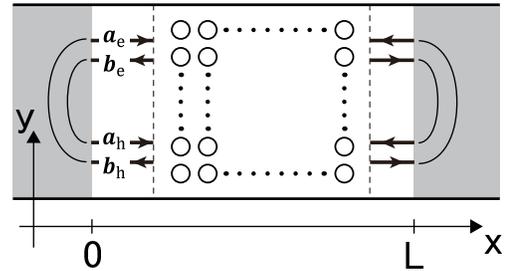}\hspace{4mm}%
\begin{minipage}[b]{9cm}
\caption{
Model for a semiconductor nanowire Josephson junction.
A tight-binding model is applied for the normal region
with hard-wall potentials for the nanowire.
The nanowire is represented by a quasi-one dimensional
system along the $x$ direction. The normal region is
$0<x<L$ and the superconducting region
induced by the proximity effect is $x<0$, $x>L$.
}
\end{minipage}
\end{figure}

The nanowire along the $x$ direction is connected to
two superconductors (Fig.\ 1).
At $x<0$ and $x>L$, the superconducting pair potential is
induced into the nanowire by the proximity effect.
We assume that the pair potential is
$\Delta (x) =\Delta_0 e^{i\varphi /2}$ at $x<0$ and
$\Delta (x) =\Delta_0 e^{-i\varphi /2}$ at $x>L$, where
$\varphi$ is the phase difference between the two superconductors.
In the normal region at $0<x<L$, $\Delta (x) =0$.
When a magnetic field is applied to the junction,
the Zeeman effect is taken into account in the nanowire.
The magnetic field is not too large to break the superconductivity
and  screened in the superconducting regions.
The Hamiltonian is given by $H = H_0 + H_{\rm SO} + H_{\rm Z}$
with $H_0 = \bm{p}^2 /2m + V_{\rm conf} + V_{\rm imp}$,
the Rashba interaction
$H_{\rm SO} = (\alpha /\hbar) (p_y \sigma_x - p_x \sigma_y)$,
and the Zeeman term
$H_{\rm Z} = g \mu_{\rm B} {\bm B} \cdot \hat{\bm{\sigma}}/2$,
using effective mass $m$, $g$-factor $g$ ($< 0$ for InSb),
Bohr magneton $\mu_{\rm B}$, and Pauli matrices $\hat{\bm{\sigma}}$.
We neglect the orbital magnetization effect in the nanowire.
$V_{\rm conf}$ describes the confining potential for
the nanowire. $V_{\rm imp}$ represents the impurity potentials.
We consider short junction, where the spacing between
two superconductors is much smaller than the coherent length
in the normal region, $L \ll \xi$. There is no potential barrier at
$x=0,L$. The Zeeman energy $E_{\rm Z} = |g \mu_B B|$ and
the pair potential $\Delta_0$ are much smaller than
the Fermi energy $E_{\rm F}$.

The Bogoliubov-de Gennes (BdG) equation is written as
\begin{equation}
\left( \begin{array}{cc}
H - E_{\rm F} & \hat{\Delta} \\
\hat{\Delta}^\dagger & -(H^* - E_{\rm F})
\end{array} \right)
\left( \begin{array}{c}
\bm{\psi}_{\rm e} \\
\bm{\psi}_{\rm h}
\end{array} \right)
= E \left( \begin{array}{c}
\bm{\psi}_{\rm e} \\
\bm{\psi}_{\rm h}
\end{array} \right)
\label{eq:BdG}
\end{equation}
with $\hat{\Delta} = \Delta (x) \hat{g}$.
$\bm{\psi}_{\rm e} = (\psi_{{\rm e} +}, \psi_{{\rm e} -} )^{\rm T}$
and $\bm{\psi}_{\rm h} = (\psi_{{\rm h} +}, \psi_{{\rm h} -} )^{\rm T}$
are the spinors for electron and hole, respectively.
$\hat{g}=-i \hat{\sigma}_y$.
The energy $E$ is measured from the Fermi level $E_{\rm F}$.
The BdG equation determines the  Andreev levels $E_n$
($|E_n|<\Delta_0$) as a function of $\varphi$.

The ground state energy of junction is given by
$E_{\rm gs} (\varphi) =-(1/2)
{\sum_n}^{\prime} E_n (\varphi )$,
where the summation is taken over all the positive
Andreev levels, $E_n (\varphi)>0$.
The contribution from continuous levels ($|E| >\Delta_0$) can
be disregarded for the short junctions~\cite{Beenakker}.
At zero temperature, the supercurrent is calculated as
$ I (\varphi) =(2e/\hbar) (d E_{\rm gs}/d \varphi)$.
The current is a periodic function for $-\pi \le \varphi < \pi$.
The maximum (or absolute value of minimum) of $I (\varphi)$
yields the critical current $I_{\rm c}$.

The BdG equation in eq.\ (\ref{eq:BdG}) is expressed in terms of
the scattering matrix~\cite{Beenakker}. The scattering matrix of
electrons (holes) transport in the normal region is given by
$\hat{S}_{\rm e}$ ($\hat{S}_{\rm h}$). $\hat{S}_{\rm e}$ and
$\hat{S}_{\rm h}$ are related to each other by
$\hat{S}_{\rm e} = \hat{S}_{\rm h}^*$ for the short junctions.
We denote $\hat{S}_{\rm e} = \hat{S}$ and $\hat{S}_{\rm h} = \hat{S}^*$.
The Andreev reflection at $x=0$ and $L$ is
described by the scattering matrix $\hat{r}_{\rm he}$
for the conversion from electron to hole
and $\hat{r}_{\rm eh}$ for that from hole to electron.
The normal reflection can be neglected.
The matrix coefficients of $\hat{r}_{\rm he}$ and $\hat{r}_{\rm eh}$,
e.g., $\exp \{ -i \arccos(E /\Delta_0) - i\varphi /2 \}$ for
$\hat{r}_{\rm he}$ at $x=0$, are calculated from
the boundary condition at $x=0$ and $L$. The SO interaction
does not affect the Andreev reflection coefficients.
The Andreev levels, $E_n (\varphi)$, are obtained from
the product of $\hat{S}$, $\hat{r}_{\rm he}$,
and $\hat{r}_{\rm eh}$,
\begin{equation}
\det \left(\hat{1} - \hat{r}_{\rm eh} \hat{S}^*
\hat{r}_{\rm he} \hat{S} \right) =0.
\label{eq:determinant}
\end{equation}
Equation (\ref{eq:determinant}) is equivalent to
the BdG equation (\ref{eq:BdG}).

To calculate the scattering matrix $\hat{S}$, we adopt the tight-binding
model which discretizes a two-dimensional space ($xy$ plane).
The edges of nanowire are represented by a hard-wall potential.
The width of nanowire is $W=12a$ with the lattice constant $a =10\mathrm{nm}$.
The Fermi wavelength is fixed at $\lambda_{\rm F} =18a$, where
the number of conduction channels is unity. The length of
normal region is $L=50a$. The on-site random potential by
impurities is taken into account, the distribution of
which potential is uniform. We set that the mean free path
due to the impurity scattering is $l_{\rm mfp}/L = 1$.
The SO length is $l_{\rm SO} /L=0.2$ with
$l_{\rm SO} = k_\alpha^{-1} =\hbar^2/(m\alpha)$.
The magnetic field is ${\bm B} = B \hat{\bm e}_\theta$ with
the angle $\theta$ from the $x$ axis in the $xy$ plane.

\section{Results}

We consider a sample for the nanowire.
For the magnetic field, we introduce a parameter,
$\theta_B = E_{\rm Z} L/(\hbar v_{\rm F})$, which
means an additional phase due to the Zeeman effect in
the propagation of electron and hole. Here, $v_{\rm F}$ is
the Fermi velocity in the absence of SO interaction.

\begin{figure}[t]
\begin{center}
\includegraphics[width=85mm]{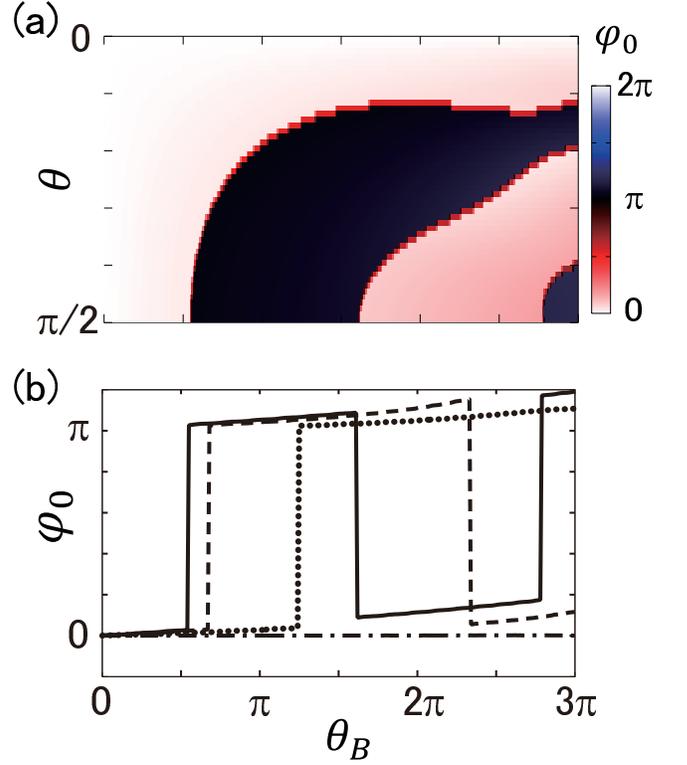}
\end{center}
\caption{
Numerical results of phase difference $\varphi_0$ at
the minimum of ground-state energy when the number of conduction
channel is unity and $l_{\rm mfp}/L=1$. The SO interaction is
$l_{\rm SO}/L =0.2$. The results are for a sample.
(a) Grayscale plot of $\varphi_0$ in the plane of magnetic field
$\theta_B = E_{\rm Z} L/(\hbar v_{\rm F})$ and its direction $\theta$.
(b) Cross section of panel (a) at $\theta =0.5\pi$ (solid),
$0.3\pi$ (broken), $0.15\pi$ (dotted), and $0$ (dotted broken lines).
}
\end{figure}

Figure 2 shows the phase difference $\varphi_0$ at the minimum of
$E_{\rm gs}$ when the magnetic field is increased and rotated.
In the absence of SO interaction, $\varphi_0$ takes only $0$ or
$\pi$ exactly and clear $0$-$\pi$ transition happens
(see Ref.\ \cite{YEN2}). In the presence of SO interaction,
$\varphi_0$ is deviated from $0$ and $\pi$, where the anomalous
Josephson current is obtained. When the magnetic field is in
the $y$ direction ($\theta =\pi/2$), the transition between
$\varphi_0 \approx 0$ and $\varphi_0 \approx \pi$ takes place
around $\theta_B = \pi/2, 3\pi/2, \cdots$.
The transition points are shifted gradually to large $\theta_B$
with decreasing of angle $\theta$. At $\theta \approx 0$,
the transition is not observed in Fig.\ 2.

The transition points in $\varphi_0$ corresponds to the position of
cusps of critical current. Figure 3(a) exhibits the critical current
when the magnetic field increases. The critical current $I_{\rm c}$
oscillates as a function of $\theta_B$. The distance of cusps is
longer when the direction of magnetic field is tilted from the $y$
axis. For the parallel magnetic field to the nanowire ($\theta =0$),
$I_{\rm c}$ has no cusp in accordance with no transition.
The critical current decreases with increase of $\theta_B$
although $\varphi_0$ is almost fixed at zero in Fig.\ 2(b).
Figure 3(b) shows $I_{\rm c}$ when the magnetic field is rotated in
the $xy$ plane. The strength of magnetic field is fixed for each line.
We find the transition by the angle $\theta$ in Fig.\ 2(a).
The critical current also oscillates as a function of $\theta$.
For small magnetic field ($\theta_B <\pi/2$), $I_{\rm c}$ changes
monotonically. The oscillation by $\theta$ is obtained for
large magnetic field. If the state of junction at $\theta =\pi/2$
is $\varphi_0 \approx \pi$ ($0$), the critical current shows
one cusp (two cusps) in Fig.\ 3(b). Therefore we can estimate
the state at $\theta =\pi/2$ from the number of cusps.

\begin{figure}[t]
\begin{center}
\includegraphics[width=80mm]{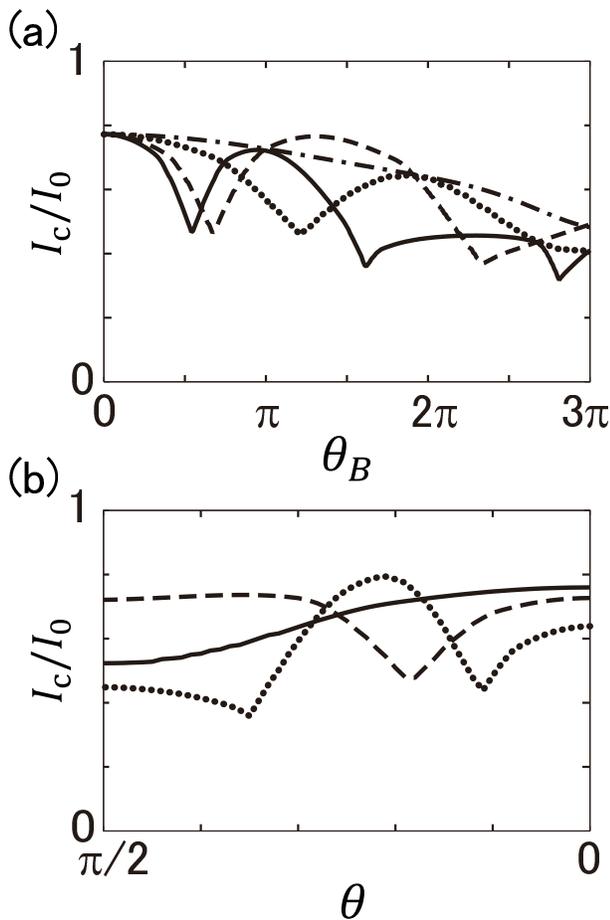}
\end{center}
\caption{
Numerical results of critical current $I_{\rm c}$ when $N=1$ and
$l_{\rm mfp}/L=1$. The SO interaction is $l_{\rm SO}/L =0.2$.
$I_0 \equiv e\Delta_0/\hbar$. The sample is the same as that in Fig.\ 2.
(a) $I_{\rm c}$ as a function of magnetic field
$\theta_B = E_{\rm Z} L/(\hbar v_{\rm F})$ when
$\theta =0.5\pi$ (solid), $0.3\pi$ (broken), $0.15\pi$ (dotted),
and $0$ (dotted broken lines).
(b) $I_{\rm c}$ as a function of magnetic field orientation $\theta$
when $\theta_B =\pi/2$ (solid), $\pi$ (broken), and $2\pi$ (dotted lines).
}
\end{figure}

\section{Conclusions and Discussion}

We have studied the DC Josephson effect in the semiconductor
nanowire with strong SO interaction. We have examined a numerical
simulation using the tight-binding model in the case of short junction.
The combination of SO interaction and Zeeman effect in the nanowire
results in the anomalous Josephson effect. The critical current oscillates
as a function of magnetic field. In the presence of SO interaction,
the oscillation of critical current depends on the magnetic field orientation.
The oscillation period is shorter when the magnetic field is perpendicular
to the nanowire. For a parallel magnetic field to the nanowire,
the transition and the cusp of critical current are not found.

In this numerical model, we have considered the Rashba interaction.
In the quasi-one-dimensional nanowire, the effective magnetic
field induced by the Rashba interaction is in the $y$ direction.
The anisotropy of critical current oscillation is understood
intuitively by the spin precession in the propagation of electron
and hole. When the external magnetic field is parallel to the effective field
(the $y$ direction), the spin quantization axis is fixed in that direction.
The electron and hole receive the additional phase in the propagation.
On the other hand, when the external field is in the $x$ direction,
the spin quantization axes for electron and hole are not parallel with
each other since the effective fields for electron and hole are
antiparallel. The spin of electron and hole forming the Andreev bound
state is rotated, which rotation cancels out the phase $\theta_B$ due
to the Zeeman splitting. As a result, the critical current oscillation
disappears. In the case of general SO interaction, the effective field
would be deviated from the $y$ axis. By measuring the anisotropy of
critical current oscillation, we can evaluate the direction of effective
field due to the SO interaction.

\section*{Acknowledgments}
We acknowledge financial support by the Motizuki Fund of
Yukawa Memorial Foundation.
We acknowledge fruitful discussions about experiments with
Professor L.\ P.\ Kouwenhoven, A.\ Geresdi, V.\ Mourik, K.\ Zuo of
Delft University of Technology.
T.Y. is a JSPS Postdoctoral Fellow for Research Abroad.

\end{document}